\documentclass[RNAAS]{aastex62}

\usepackage{graphicx}   

\begin{document}

\title{Predictions for strong lens detections with the Nancy Grace Roman Space Telescope}

\correspondingauthor{Chris Sedgwick}
\email{chris@sedgwick.uk.net}

\author{Charles Weiner}
\affiliation{School of Physical Sciences, The Open University, Milton Keynes MK7 6AA, UK}

\author[0000-0002-0517-7943]{Stephen Serjeant}
\affiliation{School of Physical Sciences, The Open University, Milton Keynes MK7 6AA, UK}


\author{Chris Sedgwick}
\affiliation{School of Physical Sciences, The Open University, Milton Keynes MK7 6AA, UK}

\keywords{infrared astronomy --- 
gravitational lenses --- cosmology observations --- galaxy evolution}


\begin{abstract}
    Strong gravitational lensing is an ideal tool for mapping the distribution of dark matter and for testing the values of cosmological parameters. Forthcoming surveys such as Euclid and the Rubin Observatory LSST should increase the number of known strong lensing systems significantly, with for example over 100,000 such systems predicted in the Euclid wide survey.  In this short research note we predict that approximately 17,000 strong gravitational lenses will also be detectible in the Nancy Grace Roman Space Telescope 
    2000 square degree survey, using the LensPop gravitational lensing model.
    We present predicted distributions in source and deflector redshifts, magnitudes, and magnifications. 
    Although this 
    survey is not primarily designed as a strong lensing detection experiment, it will still provide a large complementary catalogue to shallower, wider-area forthcoming lensing discovery projects. 
\end{abstract}

\section{Introduction}\label{sec:intro}

Gravitational lensing depends only on the distribution and geometry of foreground matter, and is independent of 
its composition 
or luminosity. It is therefore an ideal tool for mapping the distribution of 
dark matter and 
testing the values of cosmological parameters. 
In addition, magnification by the lens can 
enhance the spatial resolution of background galaxies by $1-2$ orders of magnitude.  

Strong gravitational lenses have been detected in various previous and ongoing surveys. The Sloan Lens ACS Survey (SLACS; Bolton et al.\ 2006) combined the massive data volume of the Sloan Digital Sky Survey (SDSS) with the high-resolution imaging capability of the Hubble Space Telescope to identify and study some 100 lenses and lens candidates (Auger et al.\ 2009). Neural networks are increasingly 
used to identify lenses in data catalogues: for example, a recent study of the Dark Energy Survey (DES) using convolutional neural networks found 84 strong lens candidates (Jacobs et al.\ 2019). Citizen science is also being used to find lenses which are not always clear to automated searches (Marshall et al.\ 2016). These studies (and serendipitous discoveries) have brought the total number of strong lenses discovered to date to several hundred. 

Forthcoming surveys should increase this number significantly. The Large Synoptic Telescope Survey (LSST; Abell et al.\ 2009, Ivezic et al.\ 2008) will start observing in 2022. Also highly significant for the discovery of strong lenses will be two space-borne instruments:  {\it Euclid} (Laureijs et al.\ 2011), a European Space Agency project planned to launch in 2022, and the {\it Nancy Grace Roman Space Telescope}, formerly known as the Wide Field InfraRed Survey (WFIRST; Green et al.\ 2012), a NASA 
mission 
planned to launch in 2025. 

Two detailed predictions of strong lensing have already been made for 
{\it Euclid}: 
Serjeant (2014) presented selection techniques to reliably identify 
up to $10,000$
strong gravitational lenses using {\it Euclid}'s spectroscopic surveys, and the gravitational lens model LensPop developed by Collett (2015) predicted 170,000 strong lenses to be detectable 
in {\it Euclid}'s wide imaging survey.
This paper presents the first prediction of detectable strong lenses for the {\it Roman Space Telescope}, based on the LensPop model.\\\\


\section{Methodology}\label{sec:model}

The LensPop model is described in Collett (2015) with 
its source code available as open source software\footnote{https://github.com/tcollett/LensPop}. 
The model has required only minor modifications to extend it to the {\it Roman Space Telescope}.

The model assumes strong lensing by elliptical galaxies, modelled as singular isothermal ellipsoids (SIEs).  Firstly, a population of foreground SIEs is generated with five key parameters: redshift, stellar velocity dispersion, flattening, effective radius and absolute magnitude. Secondly, the galaxy population simulated for the LSST collaboration (Connolly et al.\ 2010) is assumed for background sources. The lensing cross-section of the foreground population is then projected onto the background source populations to generate an idealised set of lens systems (deflector + source). Finally, the model applies the observing parameters of the survey being considered to discover the final set of strong lenses detectable by that survey.

Four criteria are used for accepting the lens: the source is within the Einstein radius; the image is resolved; tangential shearing of arcs is detectible; the signal-to-noise ratio is over 20:

\begin{equation}\label{equation:equation1}
\theta^{2}_{E}>x^2_s+y^2_s
\end{equation}

\begin{equation}\label{equation:equation2}
\theta^{2}_{E}>r^2_s+(s/2)^2
\end{equation}

\begin{equation}\label{equation:equation3}
\mu r_s>s,   \mu>3
\end{equation}

\begin{equation}\label{equation:equation4}
{\rm SNR}>20
\end{equation}

\noindent where $\theta_{E}$ is the Einstein radius, (x$_s$, y$_s$) are the coordinates of the centre of the source with respect to the lens, r$_s$ is the source radius, s is the point spread function full width half maximum 
(seeing, in ground-based contexts), $\mu$ is the total magnification and SNR is the signal-to-noise ratio of the lensed residual.

\section{Application to the Roman Space Telescope}\label{sec:wfirst}

The {\it Roman Space Telescope} will have six filters, but our work is based on the J band (referred to by its midpoint as J\_129), which has the greatest 
AB magnitude depth of 26.9.
The survey parameters added to the code included survey area 2000 deg$^2$, 
point spread function full width half maximum of 
0.12\arcsec ~and sky brightness 23.5 mags arcsec$^{-2}$.

Another modification to the code was made to produce values for source and lens  J\_129 magnitudes. An extrapolation was made from SDSS values for i and z bands:
the wavelength difference between the i and z bands is $\simeq0.1\mu$m, and between z and J bands is $\simeq0.44 \mu$m, so a simple log-linear extrapolation gives
$m_{J}=m_{z}-(m_{i}-m_{z})*4.4$.
We use this phenomenological extrapolation over a short wavelength range to minimise the assumptions about galaxy SEDs.

The {\it Roman Space Telescope} survey is intended to cover 2000 deg$^2$ over a five year period, so each sky position is expected to have several hundred
seconds integration time. We have estimated an effective exposure time of 700 seconds in the model. For this value, we verified that a sky brightness of 23.5 mags arcsec$^{-2}$ is consistent with a 5$\sigma$ point source sensitivity of magnitude $\sim26.9$ AB in the J\_129 band in simulated lens-subtracted images. 

This work assumes $H_0$=70.0 kms$^-$$^1$Mpc$^-$$^1$, $\Omega_M$=0.3 and $\Omega_{\Lambda}$=0.7. Further details are given in Weiner (2019).

\section{Results}\label{sec:results}

\begin{figure}[h!]
\begin{center}
   \resizebox{1.6in}{!}{\includegraphics{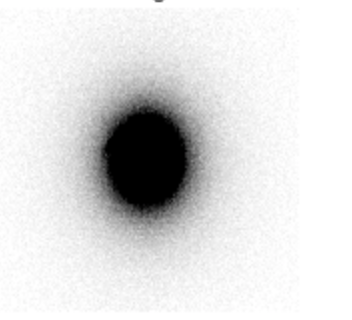}} 
   \resizebox{1.45in}{!}{\includegraphics{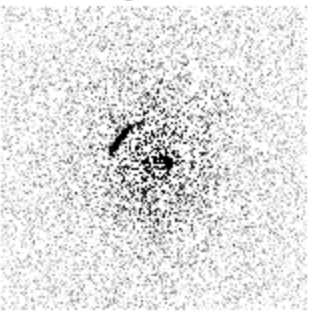}}  
   \resizebox{2.6in}{!}{\includegraphics{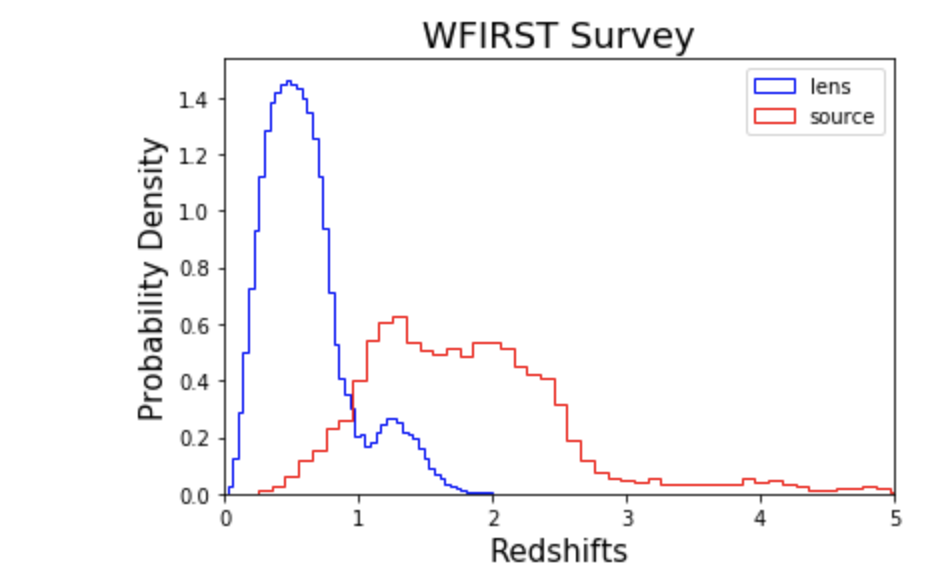}} 
    
    \resizebox{1.64in}{!}{\includegraphics{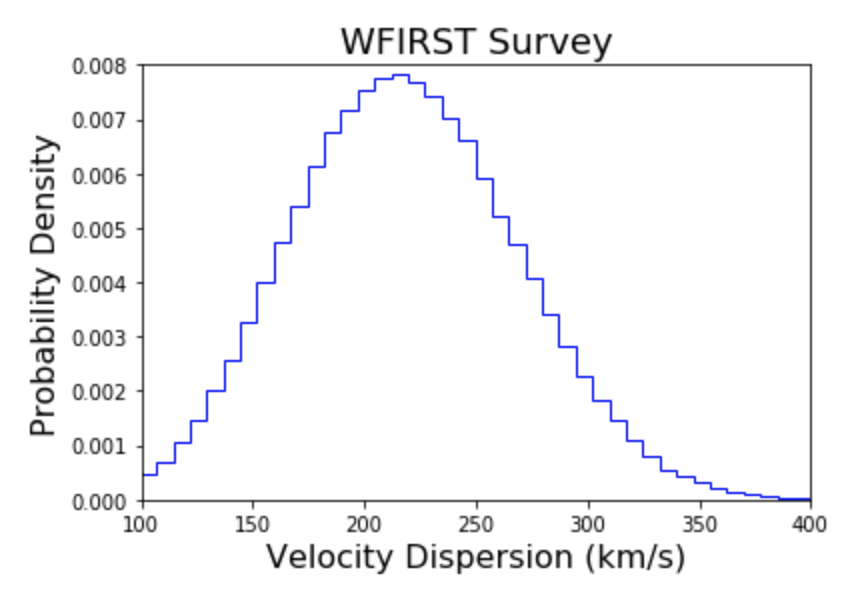}}    
    \resizebox{1.64in}{!}{\includegraphics{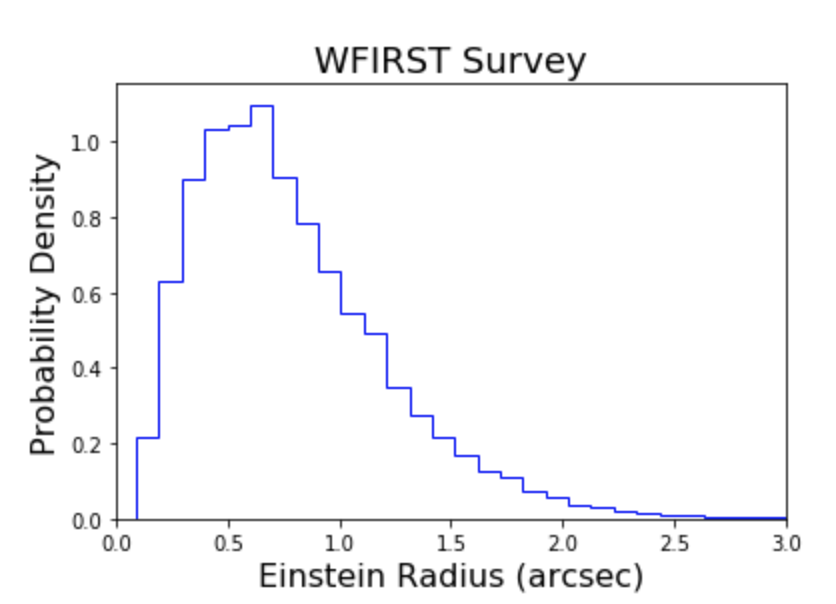}}    \resizebox{1.64in}{!}{\includegraphics{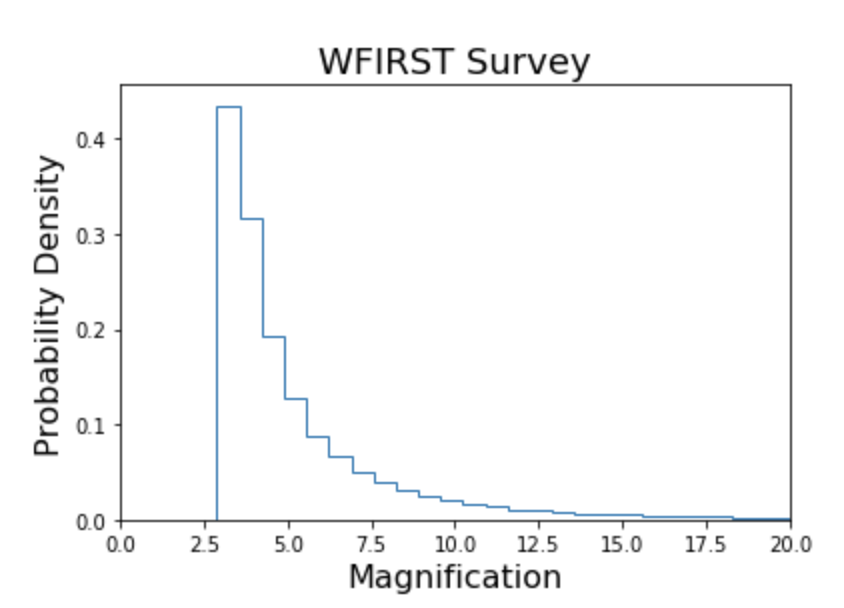}}   
    \resizebox{1.64in}{!}{\includegraphics{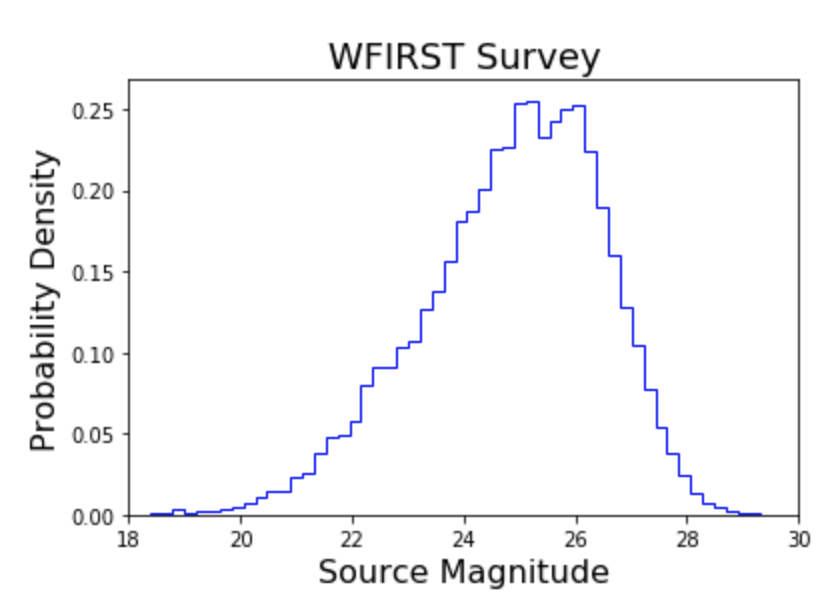}}

\caption{
Image of a simulated lens detectable by the {\it Roman Space Telescope} in the J\_129 filter band; image minus lens (assuming
subtraction is Poisson noise limited); histograms of distributions of predicted redshifts;
the velocity dispersion of the lens;
the Einstein radius; 
the distribution of magnifications; and 
source magnitudes.\label{fig:1}}
\end{center}
\end{figure}

The model predicts 16,778 detectable strong lenses for the {\it Roman Space Telescope}. Histograms of key predictions are shown in  in Figure~\ref{fig:1}: lens redshifts (mean 0.6$\pm$0.3), source redshifts (mean 1.9$\pm$0.9), magnification (mean 5.5$\pm$3.9), velocity dispersion of lens (mean 219$\pm$50), Einstein radius (mean 0.8\arcsec$\pm$0.4\arcsec) and source magnitude (mean 24.8$\pm$1.6). A typical image of a predicted lens system, and the image minus the lens are also 
in Figure~\ref{fig:1}.


The predicted average magnification 
is lower than 
for {\it Euclid} (5.5 compared to 7.2), although with a smaller variance. Predicted mean redshifts for both lenses and sources are similar in each case. For comparison, the model predicts 39,000 strong lenses will be detectable by the LSST.

\section{Discussion and summary}\label{sec:discussion}

The model we have used is based on simulated data and an assumption that SIEs account for most lens systems. For the background sources, the LSST simulated source catalogue is limited in depth, so the model may be under-predicting the number of highly magnified sources.
On the other hand, sources as faint as the limiting magnitude of 26.9 are likely to be extended sources and so may not be detected since the flux will be spread over several pixels. Actual lens discoveries will depend on the methods developed to identify such large numbers of lenses in this survey.
 
It would seem that a significant increase, potentially up to two orders of magnitude, of strong lenses available for further study will be provided by the two new space telescopes. This will provide the opportunity for a great deal of new science on dark matter, cosmological parameters and on early galaxy formation from the distant sources discovered, giving key targets for follow-up observations.

\acknowledgments
We thank Tom Collett,
the Project Scientists in the WFIRST Communications Team 
and Judith Croston.
SS thanks STFC for support 
grant ST/P000584/1.


\begin{thebibliography}{}
\bibitem[Abell et al. (2009)]{}Abell, P.A., Burke, D.L., Hamuy, M. et al.\ 2009, LSST science book v2.0, Technical report
\bibitem[Auger et al. (2009)]{2009ApJ...705.1099A}Auger, M., Treu, T., Bolton, A.\ et al.\ 2009, ApJ, 705, 1099
\bibitem[Bolton et al. (2006)]{2006ApJ...638..703B}Bolton, A.S., Burles, S., Koopmans, L.V. et al.\ 2006, ApJ, 638, 703
\bibitem[Collett (2015)]{2015ApJ...811...20C}Collett, T.E., 2015, ApJ, 811, 20
\bibitem[Connolly et al. (2010)]{2010SPIE.7738E..1OC}Connolly, A.J., Peterson, J., Jernigan, J.G. et al.\ 2010, Proc. SPIE, 7738, 7738IO
\bibitem[Green et al. (2012)]{2012arXiv1208.4012G}Green, J., Schechter, P., Baltay, C. et al.\ 2012, arXiv 1208.4012
\bibitem[Ivezic et al. (2008)]{2009AAS...21346003I}Ivezic, Z., Tyson, J., Abel, B. et al.\ 2008, arXiv 0805.2366
\bibitem[Jacobs et al. (2019)]{2019MNRAS.484.5330J}Jacobs, C., Collett, T., Glazebrook K. et al.\ 2019, MNRAS, 484, 5330
\bibitem[Laureijs et al. (2011)]{2011arXiv1110.3193L}Laureijs, R., Amiaux, J., Arduini, S. et al.\ 2011, arXiv 1110.3193
\bibitem[Marshall et al. (2016)]{2016MNRAS.455.1171M}Marshall, P.J., Verma, A., More, A.\ et al.\ 2016, MNRAS, 455, 1171
\bibitem[Serjeant (2014)]{2014ApJ...793L..10S}Serjeant, S., 2014, ApJ, 793, L10 
\bibitem[Weiner(2019)]{OU}Weiner, C.F., 2019, MPhil thesis (The Open University)

\end{thebibliography}
\end{document}